\documentclass{aastex}
\usepackage{emulateapj5}
\usepackage{apjfonts}
\usepackage{float}
\usepackage{graphics}
\newcommand{\be}{\begin{equation}}
\newcommand{\ee}{\end{equation}}
\def\ergs{{\rm\,erg\,s^{-1}}}
\newcommand{\msun}{{M}_{\sun}}

\slugcomment{ApJ in press} \shorttitle{Jet power from ADAF and the
possible applications} \shortauthors{Wu, Q. \& Cao, X. }

\begin{document}

\title{Jet power extracted from ADAF and the applications to X-ray binaries and radio galaxy FRI/FR II dichotomy}

\author{Qingwen Wu\altaffilmark{1} and Xinwu Cao\altaffilmark{2,3}}

\altaffiltext{1}{International Center for Astrophysics, Korean
Astronomy and Space Science Institute, Daejeon 305348, Republic of
Korean; qwwu@shao.ac.cn}

\altaffiltext{2}{Shanghai Astronomical Observatory, Chinese Academy
of Sciences, Shanghai, 200030  China}

\altaffiltext{3}{Joint Institute for Galaxy and Cosmology (JOINGC)
of SHAO and USTC, 80 Randan Road, Shanghai 200030, China}

\begin{abstract}
   We calculate the jet power of the classical Blandford-Znajek (BZ) model and hybrid
  model developed by Meier based on the global solutions of advection dominated accretion flows (ADAFs)
  surrounding Kerr black holes. We find that the jet
  power of the hybrid model is larger than that of the pure BZ
  model. The jet power will dominate over the accretion power, and
  the objects will enter into ``\emph{jet-power-}dominated advective systems," when
  the accretion rate is less than a critical value $\dot{m}_{\rm c}=\dot{M}_{\rm c}/\dot{M}_{\rm Edd}$,
  where $3\times10^{-4}\lesssim\dot{m}_{\rm c}\lesssim 5\times10^{-3}$ is a function of black hole spin
  parameter.
  The accretion power will be dominant when $\dot{m}\gtrsim \dot{m}_{\rm
  c}$ and the objects will enter into ``\emph{accretion-power-}dominated advective
  systems."
  This is roughly consistent with that constrained from the
  low/hard-state black hole X-ray binaries (e.g., Fender et al.).

  We calculate the maximal jet power as a function of black hole mass with
  the hybrid jet formation model, and find it can roughly reproduce the dividing
  line of the Ledlow-Owen relation for FR I/FR II dichotomy in the jet power-black hole(BH) mass
  plane ($Q_{\rm jet}-M_{\rm BH}$) if the dimensionless
  accretion rate $\dot{m}\sim0.01$ and BH spin parameter $j\sim0.9-0.99$ are adopted.
  This accretion rate $\dot{m}\sim0.01$ is consistent with that of the critical accretion rate
  for the accretion mode transition of a standard disk to an ADAF constrained from the
  state transition of X-ray binaries. Our results imply that most FR I galaxies may
  be in the ADAF accretion mode similar to the low/hard-state XRBs.

\end{abstract}


\section{Introduction}

Black hole accretion is thought to power active galactic nuclei
(AGNs) and  X-ray binaries (XRBs). Both the UV/optical bumps
observed in luminous quasars and the soft X-rays observed in the
high/soft state XRBs can be naturally interpreted as blackbody
emission from a cold, optically thick, and geometrically thin
standard disk \citep*[SSD; e.g.,][]{ss73,sm89}. A hot, optically
thin, geometrically thick advection dominated accretion flow model
has been developed in the last several decades (ADAF, or ``radiative
inefficient accretion flows''; e.g., Ichimaru 1977; Rees et al.
1982; Narayan \& Yi 1994, 1995; Abramowicz et al. 1995; see Kato et
al. 2008 and Narayan \& McClintock 2008 for reviews), which can
successfully explain most features of  the nearby low-luminosity
AGNs and low/hard-state XRBs (see Remillard \& McClintock 2006; Done
et al. 2007; Yuan 2007; Ho 2008 for recent reviews). Their spectral
energy distributions (SEDs) can be well reproduced by the ADAF
model. It was found that the hard X-ray photon indices of both XRBs
and AGNs are anti-correlated with the Eddington ratios when the
Eddington ratio is less than a critical value, while they become
positively correlated when the Eddington ratio is higher than the
critical value \citep*[][and references therein]{wg08,gc08}. These
results provide evidence for the accretion mode transition near the
critical Eddington ratio.

It is widely believed that the radio emission of both the low/hard
state XRBs \citep*[e.g.][and references therein]{fe06} and the low
luminosity AGNs \citep*[LLAGNs, e.g.,][]{wc05} comes from the jets.
However, the detailed physical mechanism for the jet formation is
still unclear for either AGNs or XRBs. The currently most favored
jet formation mechanisms include the Blandford-Znajek (BZ) process
\citep{bz77} and the Blandford-Payne (BP) process \citep{bp82}. In
the BZ process, energy and angular momentum are extracted from a
rotating black hole and transferred to a remote astrophysical load
by open magnetic field lines. In the BP process, the magnetic fields
threading the disk extract energy from the rotation of the accretion
disk itself to power the jet/outflow. \citet{me99} showed that even
if the BZ process is neglected entirely, the jet power contributed
by the field threading the disk alone will still be a function of
the black hole spin since the rotating metric contributes to the
rotation of the magnetic field. Recent magnetohydrodynamic(MHD)
simulations also showed that both the BH spin and the accretion rate
in the underlying accretion disk play important roles for the jet
formation \citep*[][]{ko00,kg04,hi04,de05,hk06}. The relative
importance of these two mechanisms was explored by many authors
\citep*[e.g.,][]{ga97,li99,me01,ca02,ne07,wang08}. Assuming the
poloidal magnetic field component at the disk surface to be of the
same order as the toroidal field component, the maximal jet power
extracted from the accretion disk (BP process) may dominate over the
maximal power extracted by the BZ process
\citep*[e.g.,][]{li99,me01,ca02,ne07}. However, \citet{re06} found
that the dynamics of the accretion disk in plunging region within
the innermost stable circular orbit can greatly enhance the trapping
of large scale magnetic field on the BH, and therefore increase the
importance of the BZ mechanism effect compared to previous estimates
that ignore the plunge region \citep*[e.g.][]{ga97}.

Radio galaxies are usually classified as FR I or FR II sources
depending on their radio morphology. FR I radio galaxies (defined by
edge-darkened radio structure) have lower radio power than FR II
galaxies (defined by edge-brightened radio structure due to compact
jet terminating hot spots; Fanaroff \& Riley 1974). What causes the
morphological difference between FR I and FR II radio galaxies is
still unclear. The theoretical models fall into two different
groups: (1) the morphological differences arise of the different
physical conditions in their ambient medium \citep*[see][for a
summary]{gk00}; (2) their intrinsic difference of their central
engines, i.e., different accretion modes and/or jet formation
processes \citep*[e.g.,][]{bi95,re96a,me99,gc01,ma04,wo07,ha07}.

Most previous works on the jet power extracted from ADAFs were based
on self-similar solutions of ADAFs
\citep*[e.g.,][]{an99,me01,cr04,ne07}. The self-similar solution can
reproduce the global solution quite well at large radii, while it
deviates significantly near the black hole \citep*[e.g.,][]{na97},
where the relativistic jets are supposed to be formed. In this paper
we calculate the jet power, incorporating some recent MHD simulation
results, based on the global ADAF solutions surrounding Kerr black
holes. We then compare our model calculations with the accretion
power and jet power of XRBs and the radio galaxy FR I/FR II
dichotomy.

\section{Accretion and jet models}

\subsection{Global ADAF model}

     We calculate the global structure of an accretion flow
     using an approach similar to that of Narayan et al.
     (1997). However, the pseudo-Kerr potential for a rotating BH given by \citet{mu02} is adopted in
   solving the equations of the accretion flow, which allows us to calculate the structure of an accretion
   flow surrounding either a spinning or a nonspinning black hole.
   The simple $\alpha$-viscosity
     ($\tau_{r\phi}=\alpha p$, and $p$ is total pressure, i.e., gas pressure plus magnetic pressure) is
    adopted and all radiative processes (synchrotron, bremsstrahlung and Compton
    scattering) are included consistently in our calculations for ADAF structure.  The advection by
    ions and electrons has been considered in the energy equation, and
    a more realistic state of accreting gas (instead of a
    polytropic index $\gamma_{\rm g}$) is employed in the calculations, which is similar
    to that used by Manmoto (2000). We solve a set of hydrodynamical equations (i.e., the radial momentum,
    angular momentum, and energy equations) for an
   ADAF, and tune the parameter $l_{\rm in}$, the specific angular
   momentum of the gas swallowed by the black hole, to let the
   solution passing smoothly through the sonic point near the black
   hole \citep*[see][for details]{na97}. We find that the derived global
   solutions for Kerr black holes can reproduce
    all the essential properties of the solutions derived in full general relativistic frame
    by \citet{man00} with error less than 10\%. The global structure of
    an ADAF surrounding a BH spinning at rate $j$ with mass $M_{\rm BH}$
    can be calculated with proper outer boundaries (e.g., Manmoto 2000), if the parameters $\dot{m}$, $\alpha$,
    $\beta$,
    and $\delta$ are specified. The parameter $j=J/GM_{\rm BH}^{2}c^{-1}$ is the dimensionless angular
     momentum, where $J$ is the angular momentum of the BH, $\dot{m}=\dot{M}/\dot{M}_{\rm
    Edd}$ is dimensionless accretion rate, and  $\dot{M}_{\rm
    Edd}$ is the Eddington accretion rate defined as $\dot{M}_{\rm Edd}=1.4\times10^{18}M_{\rm BH}/\msun \rm\ g
  \  s^{-1}$.  The value of
    $\alpha$ adopted in ADAF modeling is supposed to be within a very
    narrow range, i.e., $\alpha=0.1-0.3$ \citep*[e.g.,][and references
    therein]{nm08}, which is supported by MHD numerical simulations
    of accretion flows $\sim0.05-0.2$ \citep{hb02} and the observationally-determined values $\sim0.1-0.4$
    based primarily on studies of outbursts in dwarf novae and X-ray
    transients\citep{ki07}. The magnetic parameter
   $\beta$ (defined as ratio of gas to magnetic pressure in the accretion
   flow, $\beta=P_{\rm g}/P_{\rm m}$) is not an independent
   parameter and can be related to $\alpha$ as
   $\beta\simeq(0.55-\alpha)/\alpha$, as suggested by MHD
   simulations \citep*[e.g.,][]{ha95}, where $P_{\rm m}=B_{\rm dynamo}^{2}/8\pi$
     and $B_{\rm dynamo}$ is the magnetic field strength of the ADAF in the local reference frame. The parameter
   $\beta\simeq1-5$ for the typical value of $\alpha\simeq0.1-0.3$.
      Another poorly constrained parameter is $\delta$, which describes the
fraction of the turbulent dissipation that directly heats the
electrons in the flow. Recent ADAF models typically assume
$\delta\simeq0.3-0.5$ \citep*[e.g.,][see also Sharma et al. 2007 for
slightly lower $\delta$ value]{yu03,wu07}.

Two important modifications of the above global ADAF model are also
included \citep*[see][for more details]{me01,ne07}. First, as viewed
from an outside observer at infinity in the Boyer-Lindquist
reference frame, the disk angular velocity $\Omega$ is a sum of its
angular velocity relative to the local metric $\Omega^{'}$ plus the
angular velocity of the metric itself in the Boyer-Lindquist frame
$\omega\equiv-g_{\phi t}/g_{\phi\phi }$, i.e.,
$\Omega=\Omega^{'}+\omega$. Second, we also take into account the
field-enhancing shear caused by frame dragging in the Kerr metric,
as first suggested by \citet{me99}, which seems to be supported by
MHD simulations \citep*[e.g.,][]{hk06}.  Following the work of
\citet{me01}, the amplified magnetic field related to the magnetic
field produced by the dynamo process in the ADAF can be expressed as
$B=gB_{\rm dynamo}$, where $g=\Omega/\Omega^{'}$ is the
field-enhancing factor.

\subsection{Evaluating the jet power of Blandford-Znajek model}

For a black hole of mass $M_{\rm BH}$ and dimensionless angular
momentum $j$, with magnetic fields $B_{\perp}$ normal to the horizon
at $R_{\rm H}=[1+(1-j^{2})^{1/2}]R_{\rm g}$ ($R_{\rm g }=GM_{\rm
bh}/c^{2}$ is the gravitational radius), the power extracted with
the BZ mechanism is given by \citep*[e.g.,][]{ga97,mt82}
 \be
  Q_{\rm jet}^{\rm BZ}=\frac{1}{32}\omega_{\rm F}^{2}B_{\perp}^{2}R_{\rm
  H}^{2}j^{2}c,
  \ee
where $\omega_{\rm F}^{2}\equiv\Omega_{\rm F}(\Omega_{\rm
H}-\Omega_{\rm F})/\Omega_{\rm H}^{2}$ is determined by the angular
velocity of field lines $\Omega_{\rm F}$ relative to that of the
hole $\Omega_{\rm H}$. In order to estimate the maximal power
extracted from a spinning BH, $\omega_{\rm F}$ is always required to
be 1/2 \citep*[e.g.,][]{li99,ca02}. The magnetic field $B_{\perp}$
is assumed to approximate to the poloidal component $B_{\rm p}$, and
\citet{li99} proposed that $B_{\rm p}\simeq B_{\rm dynamo}$ due to
the hot thick disk of ADAF ($H\sim R$). Therefore, we use
$B_{\perp}\simeq B_{\rm p}\simeq gB_{\rm dynamo}$ in our
calculations considering the field enhancing effect. This is
consistent with recent MHD simulations in which the poloidal fields
are dominant in the $z$-direction near the BH \citep{ka04}.
Following the work of \citet{ne07}, all the physical quantities are
evaluated at $R=R_{\rm ms}$.

\subsection{Evaluating the jet power of hybrid model}

As pointed out by \citet{me99}, the differential dragging of the
frames will also act as a dynamo to amplify the magnetic field, and
therefore even if the BZ process is neglected entirely, the jet
power contributed by the field threading the disk alone will still
be a function of the BH spin, since the metric of the rotating black
hole contributes to the rotation of the magnetic field. Both the BP
and BZ mechanisms were incorporated in this hybrid jet formation
model, in which the magnetic fields extract energy both from the
accretion flow and the spinning hole \citep{me99,me01,pc90}. The
total jet power for the hybrid model is given by
 \be Q_{\rm
jet}^{\rm disk}=B_{\rm p}^{2}R^{4}\Omega^{2}/32c, \ee where $R$ is
the characteristic size of jet formation region, $B_{\rm p}\simeq
gB_{\rm dynamo}$, $\Omega=\Omega^{'}+\omega$, and all quantities are
evaluated at $R=R_{\rm ms}$ \citep*[see][for details]{me01}.

\section{Results and Discussion}

     To estimate the jet power extracted from the inner region of the disk, we
    employ the global ADAF solution for a spinning black hole, and the field-enhancing shear in
    the Kerr metric has been taken into account \citep*[e.g.,][]{me01}. Our global
    calculations show that the magnetic field can be amplified
    roughly 2 times when $j\sim0.9$ in the plunging region (the zone
     between $R_{\rm ms}$ and $R_{\rm H}$). We find that
     the jet power for either the BZ model or the hybrid model is roughly
   proportional to the accretion rate/BH mass, but its dependence on $j$ is rather
   complicated. Figure 1 shows the spin dependence
    of jet power for the BZ model and the hybrid
   model for given $M_{\rm BH}=10^{8}\msun$ and
   $\dot{m}=0.01$ (also see similar results in Fig. 1 of Nemmen et al. 2007).
   The black-solid lines and blue-dashed lines in Fig. 1 denote the jet
   power for two different values of the viscosity parameter
   $\alpha=0.3$ ($\beta\simeq1$) and $\alpha=0.1$ ($\beta\simeq5$) for the case of
   $\delta=0.5$, respectively. We find that the jet power varies little for
   different viscosity parameters, provided all other parameters are fixed. The physical
   reason is that the jet power (BZ/hybrid models) $Q_{\rm jet}\propto B^{2}\propto 1/\alpha(1+\beta)\simeq
   constant$ in the ADAF \citep*[see][and relation of $\alpha$ and $\beta$ in Sect. 2.1]{ny95}.
    The red dotted lines denote the jet power for the cases of $\alpha=0.3$
    ($\beta\simeq1$, Fig. 1) and $\delta=0.1$, which indicates that the
    jet power is also not sensitive to the value of $\delta$.
    We find that the ratio of the gas
     pressure to the magnetic pressure $\sim0.4-1$ in the plunging
     region varying with different BH spin parameters $j$ is mainly due to the
field-enhancing shear
     caused by frame-dragging. Our results for the case of $\alpha=0.3$ are roughly
     consistent with the MHD simulations in Kerr metric indicating the ratio $\sim0.3-1$
     at the inner boundary \citep*[][]{de03,hi04}. Therefore, $\alpha=0.3$
     and $\delta=0.5$ are adopted in the following calculations. We find that the values of these parameters
     will not affect our main conclusions.

   We find that the jet power of the hybrid model, $Q_{\rm jet}^{\rm
   disk}$, is nearly 25 times larger than that of the BZ model, $Q_{\rm jet}^{\rm
    BZ}$, for a rapidly spinning black hole with $j\sim0.9$, or even higher for a smaller $j$ (Fig. 1).
    The jet efficiency [defined as $\eta_{\rm jet}=Q_{\rm jet}/\dot{M}(R_{\rm
ms})c^{2}$]
    for a spinning BH with $j=0.99$ is 0.8\% and 20\% for the BZ model
    and the hybrid model, respectively. This
    suggests that the hybrid model plays a more important role than
    the pure BZ model in the jet formation for a black hole surrounded by an ADAF. Our
    results of the hybrid model are consistent with  $\eta_{\rm jet}\sim21\%$ for
    $j=0.99$ in the numerical MHD
    simulations \citep{hk06}. Hereafter, we implicitly consider the jet power
$Q_{\rm jet}\simeq Q_{\rm jet}^{\rm
   disk}$ in our calculations, since the jet power of the BZ model is always much lower than that
   of the hybrid model for ADAFs.

\subsection{Accretion Power/Jet Power of ADAF and Comparison with XRBs}

\figurenum{1}
\centerline{\includegraphics[angle=0,width=9.0cm]{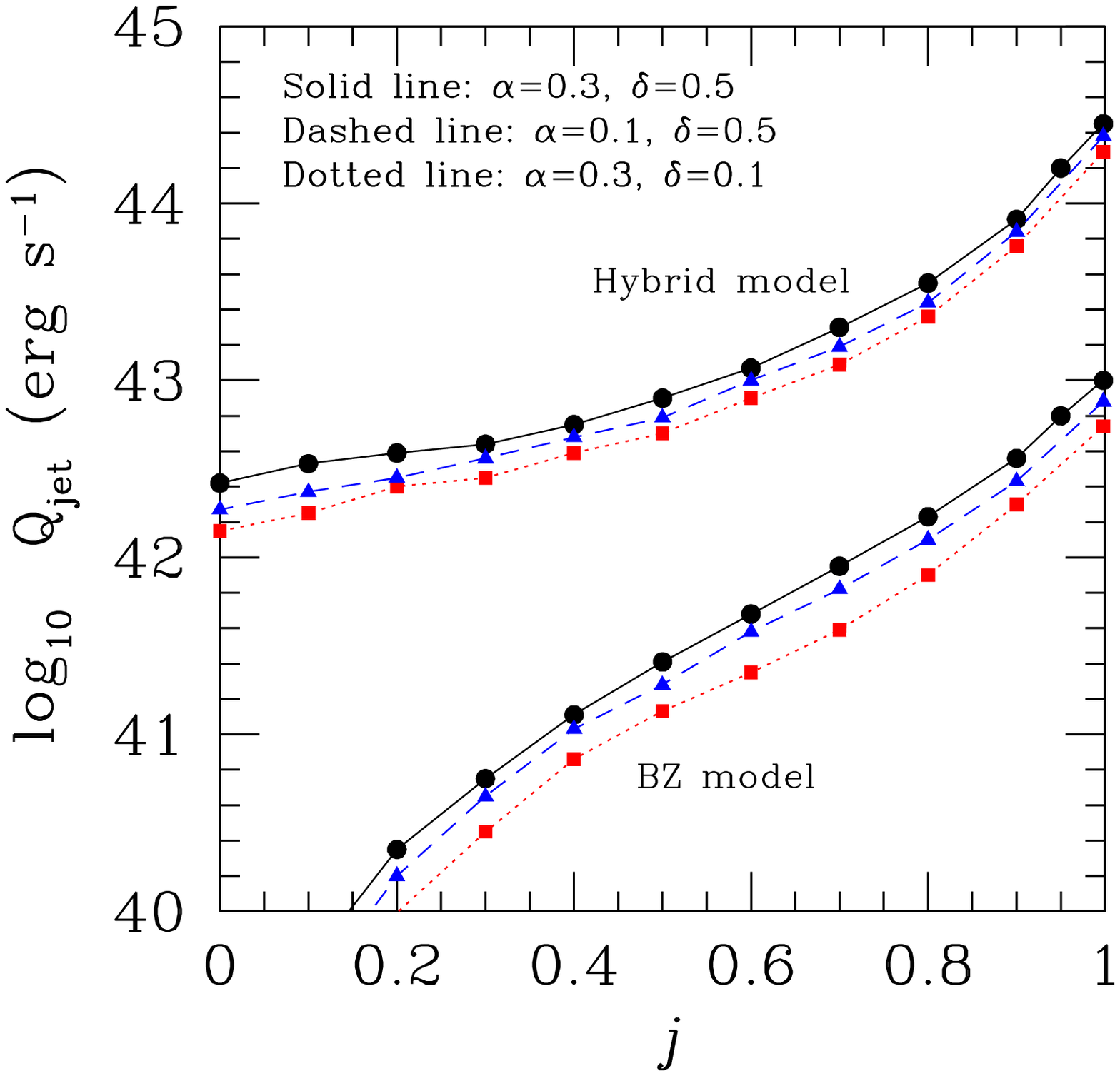}}
\figcaption{\footnotesize The relation between the jet power $Q_{\rm
jet}$ extracted from the underlying ADAF and BH spin $j$ for the BZ
and hybrid models, respectively. The black solid lines and blue
dashed lines denote $\alpha=0.3$ ($\beta=1$) and $\alpha=0.1$
($\beta=5$ and $\delta=0.5$), respectively.  The red dotted lines
denote $\alpha=0.3$ for the case of $\delta=0.1$. The BH mass
$M_{\rm BH}=10^{8}\msun$ and accretion rate $\dot{m}=0.01$ are
adopted in the calculations. \label{fig1}} \centerline{}

   Once accretion rate $\dot{m}$ falls below a critical value $\dot{m}_{\rm tr}$, the standard disk will
   transit to an ADAF at radii less than the transition radius.
   In the ADAF scenario, the accretion power (or radiated luminosity) is much lower than
   that of SSD as a result of reduced radiative efficiency. The
   power created by accreting matter can either be radiated away,
   advected into the BH, or taken away by jet/wind. Normally, the radiative efficiency of
   the pure ADAF is a function of accretion rate, i.e., $\eta=\eta(\dot{m}/\dot{m}_{\rm
   tr})^{\xi}$. The radiative efficiency $\eta$ of the ADAF is roughly proportion to the accretion rate (e.g., $\xi\sim1$)
   at lower accretion rate and slightly flattens ($\xi\lesssim1$) at near the critical value $\dot{m}_{\rm
   tr}$ \citep{mhd03,yc05,wc06,sh07}. However, the radiative
   efficiency should be slightly higher when the accretion rate is near $\dot{m}_{\rm tr}$,
   especially considering the contribution of the outer SSD which
   should be important when the transition radius is small, and/or the possibility of a condensation-feed inner
   disk \citep{liu07}. Here, we do not consider these
   two mechanisms in details due to both of them are still quite
   unclear. We assume that
   $\xi=1$ when $\dot{m}\leqslant \dot{m}_{\rm
   tr}$ (see also Narayan \& McClintock 2008), which is consistent with that constrained from the XRBs
   \citep*[e.g.,][]{ko06}, and will not affect our conclusion. Therefore, the bolometric
    luminosity (or accretion power) of the ADAF can be described by
   \be
   L_{\rm bol}=\eta(j)\frac{\dot{m}^{2}}{\dot{m}_{\rm tr}}\dot{M}_{\rm
   Edd}c^{2},
   \ee
where $\eta(j)$ is radiative efficiency depending on the BH spin
parameter $j$. The exact value of the critical accretion rate
$\dot{m}_{\rm tr}$ is still unclear. \citet{mac03} found
$\dot{m}_{\rm tr}\sim0.02$ from a detailed investigation of state
transitions in X-ray binaries. \citet{wg08} suggested that the
accretion rate for the disk transition may be 2-3 times lower than
that derived from the state transition. We adopt $\dot{m}_{\rm
tr}=0.01$ in our calculations. Figure 2 shows the relation of
$Q_{\rm jet}^{\rm disk}/L_{\rm Edd}-\dot{m}$ (\emph{solid line}) and
$L_{\rm bol}/L_{\rm Edd}-\dot{m}$ (\emph{dashed line}) for
$j=0.99,0.9,0.7$, and 0, respectively. We find that the jet power
dominates over the accretion power when the accretion rate is less
than a critical value $\dot{m}$ $\lesssim\dot{m}_{\rm c}$ and the
objects enter into the ``\emph{jet-power-}dominated advective
systems," where $3\times10^{-4}\lesssim\dot{m}_{\rm c}\lesssim
5\times10^{-3}$ depends on the BH spin parameter $j$.  The accretion
power is dominant when $\dot{m}\gtrsim \dot{m}_{\rm c}$, and the
objects are therefore ``\emph{accretion-power-}dominated advective
systems." We note that the uncertainties of $\delta$ parameter will
lead to slightly different critical accretion rate $\dot{m}_{\rm c}$
due to the radiative efficiency of ADAF, even the jet power is not
sensitive to the parameter $\delta$. We find the uncertainties on
the critical accretion rate $\dot{m}_{\rm c}$ will be less than 5
times when considering $\xi\simeq 1\pm0.2$ for
$0.1\lesssim\delta\lesssim0.5$ and BH spin $j=0-1$, and will less
than 2 times for the typical value $j=0.7-1$ found in XRBs
\citep*[e.g.,][and references therein]{ljf08}, which will not affect
our main conclusion.

\figurenum{2}
\centerline{\includegraphics[angle=0,width=9.0cm]{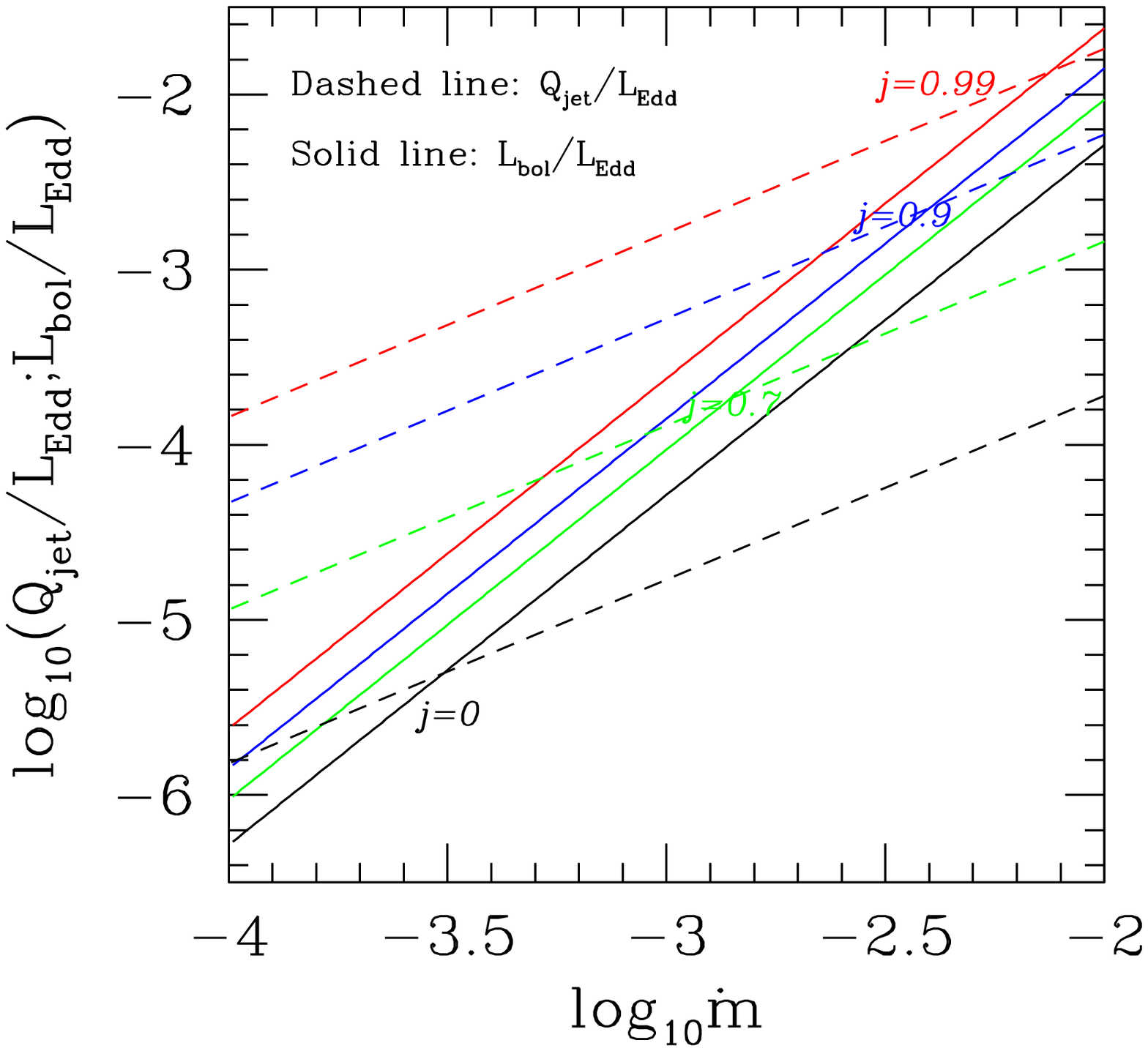}}
\figcaption{\footnotesize The relation between $Q_{\rm jet}/L_{\rm
Edd}-\dot{m}$ (dashed line) and  $L_{\rm bol}/L_{\rm Edd}-\dot{m}$
(solid line) for BH spin parameters: $j=0.99,0.9,0.7,0$(from top to
bottom), respectively. \label{fig2}} \centerline{}

  Our calculation show that the jet power $Q_{\rm jet}\propto B^{2}\propto
  \dot{m}$ and accretion power $L_{\rm bol}\propto\dot{m}^{2}$.
  The relation $Q_{\rm jet}\propto L_{\rm bol}^{0.5}$
  is consistent with the nonlinear correlation between radio
   luminosity and X-ray luminosity, $L_{\rm radio}\propto L_{\rm X}^{0.7}$,
   in the low/hard state \citep*[e.g.,][]{cor03}, when considering the simple
   optically thick conical jet model, where $Q_{\rm jet}\propto L_{\rm
   radio}^{12/17}$ \citep*[e.g.,][]{bk79,fb95}
   and $L_{\rm X}$ is an indicator of $L_{\rm bol}$.
   \citet{fe03} gave a conservative lower limit of the critical accretion rate
   $\dot{m}_{\rm c}\sim10^{-4}$ for the transition of ``accretion-power-dominated regime" and
``jet-power-dominated regime" assuming the power created by the
accreting matter is radiated (accretion power) and taken away by jet
(jet power) based on the XRB XTE J1118+480. \citet{mf06} further
suggested that these two regime transitions may occur at the
slightly higher critical value $\dot{m}_{\rm c}\sim10^{-2}$, using a
slightly higher jet power normalization. These estimates based on
low/hard state XRBs are roughly consistent with our result that the
critical accretion rate is $3\times10^{-4}\lesssim\dot{m}_{\rm
c}\lesssim 5\times10^{-3}$ when considering the uncertainties in the
estimates of the jet power normalization from the observation. We
should note that the \emph{energy advection} play an important role
for jet formation, since only a small fraction of the gravitational
energy of the flow is dissipated locally, and most of the energy
heats the protons/electrons and is advected inward. This leads to a
hot thick disk, which allows high poloidal magnetic field strength
in the inner region of the flow. We find that, as an example, the
fraction of output energy $\sim(Q_{\rm jet}+L_{\rm
Bol})/0.1\dot{M}c^{2}\sim0.2$ for $\dot{m}=10^{-3}$ and $j=0.7$, and
the other 80\% of the energy is advected into the central BH.
Therefore, the BH central engine will be a
``\emph{jet-power-}dominated advective system" when the luminosity
(or accretion rate) is less than a critical value, which is
consistent with that constrained with the observations on the BH
XRBs and neutron stars \citep*[e.g.,][]{ko06}. It should be noted
that we use ``jet-\emph{power}-dominated" not ``jet-dominated" to
discriminate the possible confusion about the origin of
multi-wavelength emission. For example, the X-ray emission in the
low/hard state XRBs and LLAGNs is still controversial, and it may be
dominated by the jet emission \citep*[e.g.,][]{mark03,fa04}, by the
underlying ADAF, or both \citep*[e.g.,][]{yc05,wu07}.

\subsection{FR I/II Dichotomy}

The division between FR I and FR II radio galaxies is clearly shown
by a line in the plane of the total radio luminosity and the optical
luminosity of the host galaxy \citep{lo96}. \citet{gc01} used the
optical luminosity of the host galaxy and the radio luminosity to
estimate the mass of its central BH and jet power, respectively.
They proposed that the FR I/FR II separation can be interpreted by
the systematically different ratios of the jet power to BH mass for
FR I and FR II sources. This implies that the FR I/FR II division is
linked to the physics (accretion and/or jet processes) on very small
scales. They argued that if the jet power is related with accretion
power, the accretion mode for low-power FR I sources may be
different from that for powerful FR II sources. There is growing
evidence to suggest that most FR I type radio galaxy nuclei may
possess ADAFs \citep*[][]{re96b,gl03,mhd03,do04,wu07}. \citet{cr04}
found that the jet power extracted from the underlying ADAFs with
pure BZ mechanism is insufficient for some high power-FR I radio
galaxies. In this work, we compare the jet power of the hybrid model
with the observed FR I/FR II dichotomy in the $M_{\rm BH}-Q_{\rm
jet}$ plane.

The dividing line between FR I and FR II sources of \citet{lo96} is
given approximately by \citet{me99} as
 \be \log P_{\rm rad}=-0.66
M_{R}+10.35, \ee where $P_{\rm rad}$ is the observed radio power at
1.4GHz (in $\rm W\ Hz^{-1}$) and $M_{R}$ is the absolute optical
$R$-band magnitude of the host galaxy. An empirical relation between
$M_{R}$ of the host galaxy and central BH mass, \be \log(M_{\rm
BH}/\msun)=-0.50(\pm0.02)M_{R}-2.96(\pm0.48), \ee was derived by
\citet{md02}. The jet power is usually estimated from the radio
luminosity by using the relation \be Q_{\rm
jet}\simeq3\times10^{38}f^{3/2}L_{151}^{6/7} \ \rm W, \ee where
$L_{151}$ is the total radio luminosity at 151 MHz in units of
$10^{28}\rm W\ Hz^{-1}\ sr^{-1}$ and the factor $f$ parameterizes
the uncertainties of the normalization which is constrained to be
between 1 and 20 \citep*[see][for details]{wi99}. \citet{br00}
argued $f$ to be most likely in the range of 10-20. The radio
luminosity at 1.4 GHz is converted to the luminosity at 151 MHz by
assuming a radio spectra index $\alpha=0.8$, where $L(\nu)\propto
\nu^{-\alpha}$. Thus, we can obtain the dividing line between jet
power and BH mass by using the Eqs. (4)-(6), \be\log Q_{\rm
jet}(\ergs)=1.13\log M_{\rm BH}(\msun)+33.42+1.50\log f. \ee

\figurenum{3}
\centerline{\includegraphics[angle=0,width=9.0cm]{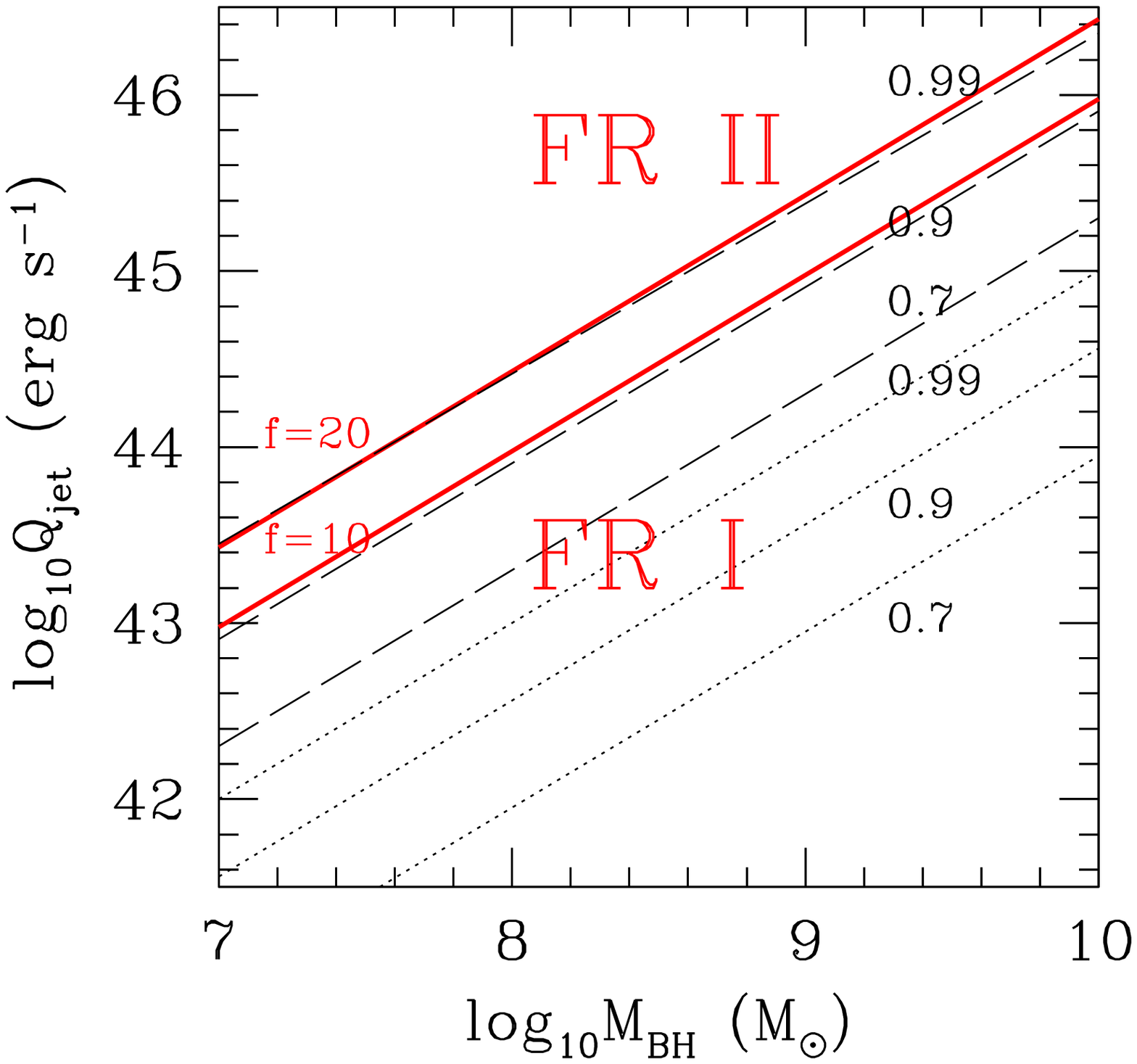}}
\figcaption{\footnotesize Jet power and BH mass dividing line
between FR I and FR II radio galaxies for the \citet{lo96} sample
(\emph{solid lines}; the jet power is calculated from the 151 MHz
radio luminosity with two different correction factors $f=20$ (top)
and $f=10$ (bottom), respectively). The dashed and dotted lines are
the jet power of the hybrid model and BZ model extracted from the
underlying ADAFs with $j=0.99,0.9,0.7$ (\emph{from top to bottom}),
respectively. The accretion rate $\dot{m}=0.01$ is adopted in all
calculations, which is roughly consistent with that constrained from
the ionization luminosity of the dividing line between FR I and FR
II sources (see the text for more details). \label{fig3}}
\centerline{}

The maximal jet power of the hybrid model (\emph{solid line}) and
the BZ model (\emph{dotted line}) for ADAFs are plotted in Fig. 3
with different black hole spin parameters: $j=0.99,0.9,0.7$,
respectively. The accretion rate $\dot{m}=\dot{m}_{\rm tr}=0.01$ is
adopted in our calculations. We find that the maximal power of the
pure BZ model is nearly 25 times less than the dividing line between
FR I and FR II radio galaxies. However, we find that the maximal jet
power of the hybrid model with $j=0.9/0.99$ can roughly reproduce
the jet power dividing line with $f=10/20$ (see Fig. 3). Thus, the
dividing line between FR I and FR II sources corresponds to the
maximal jet power of the hybrid model extracted from the ADAFs
surrounding rapidly spinning black holes accreting at the critical
rate $\dot{m}_{\rm tr}=0.01$, which is consistent with the analogue
between FR I sources and the low/hard state of XRBs
\citep*[e.g.,][]{mhd03,fa04}. We find that the jet power of the
hybrid model is very sensitive to the BH spin, especially when
$j>0.9$ (Fig. 1). Therefore, it seems that this sensitivity would
likely blur the FRI/FR II dividing line if the spins of black holes
spread around $0.9$. The FR I/FR II dividing line is indeed not very
clear (one can find a few FR I sources above the dividing line, and
vice versa for FRII sources, e.g.,  \citet{lo96}). The critical
accretion rate $\dot{m}_{\rm tr}=0.01$ for the accretion mode
transition is also supported by the ionization luminosity for the
separation between FR I and FR II sources. \citet{wi99} found that
the ionization luminosity of radio galaxies is roughly equal to the
jet power for $f=20$, which corresponds to $L_{\rm ion}/L_{\rm
Edd}\sim2.5\times10^{-2}$ for typical BH mass $M_{\rm
BH}\sim10^{7.5-9.5}\msun$ in sample of \citet{lo96}. This ionization
luminosity Eddington ratio roughly corresponds to
$\dot{m}\simeq0.01$\footnote{We note that the differences of the
Eddington ratio in \citet{gc01} and \citet{wo07} from ours are
caused by using different factor $f$, and $f=1$ is adopted in their
work, and some coefficients of Eqs. in \citet{gc01} may be erroneous
as pointed out by \citet{wo07}.}  if we assume that the bolometric
luminosity is equal to the ionization luminosity and the BH is
spinning rapidly [e.g., $\eta(j)\gtrsim0.2$ for $j\gtrsim0.95$]. It
is still unclear why the jet power of FR II radio sources is always
above this dividing line, which is beyond the scope of this work.

 Our calculations on the jet power for either the BZ or BP
mechanisms are based on the pure ADAF model. \citet{ny94} found that
the ADAF has a positive Bernoulli parameter, and the accretion flow
is therefore gravitationally unbound, which implies that the gas may
escape as outflows. In this case, the accretion rate of the
advection-dominated inflow-outflow solution (ADIOS) is a function of
the radius instead of a constant accretion rate for the pure ADAF
\citep*[e.g.,][]{bb99}. For ADIOS, the gas swallowed by the BH is
only a fraction of the rate at which it is supplied, as part of the
gas is carried away in the winds before it reaches the BH.
Therefore, both the accretion power and jet power will decrease in
the ADIOS compared the pure ADAF case for given accretion rate at
the outer boundary. The accretion power $L_{\rm bol}$ is roughly
$\propto\dot{m}^{2}_{\rm in}$, while the jet power $Q_{\rm jet}$ is
roughly $\propto\dot{m}_{\rm in}$ at the inner edge of the flow. It
means that, in Fig. 2, the lines for $L_{\rm bol}$ shift down more
than those for $Q_{\rm jet}$ in the presence of winds. Thus, the
critical accretion rate $\dot{m}_{\rm c}$ for the transition between
``\emph{jet-power-}dominated advective systems" and
``\emph{accretion-power-}dominated advective systems" in the ADIOS
should be slightly higher than that derived from the pure ADAF case.
Therefore, the critical accretion rate in the ADIOS maybe close to
$\dot{m}_{\rm c}\sim10^{-2}$ for the case of $j\gtrsim0.9$, which is
roughly consistent with that constrained from the XRBs
\citep*[e.g.,][]{mf06}. The ADIOS has a structure and a upper limit
on the accretion rate at the inner edge of the disk similar to those
of a pure ADAF if the wind is not very strong
\citep*[e.g.,][]{ch02,ne07}. The jet power of BZ/BP mechanisms is
dominantly extracted from the inner region of the flow, so the
maximal jet power for an ADIOS should be similar to that for an ADAF
without winds, provided that their accretion rate at the inner edge
of the flow are the same. Therefore, the maximal jet power from
ADIOS with $\dot{m}_{\rm in}\simeq0.01$ and $j\sim0.9-0.99$ can
still reproduce the dividing line of the FR I/FR II dichotomy (Fig.
3).

\section{Summary and Conclusion}

    The main conclusions of this work can be summarized as follows:

    (1) We investigate the jet power of the BZ mechanism and hybrid
   mechanism for the ADAFs surrounding rotating black holes based on our
   global ADAF solutions in Kerr metric. We find that the jet power of
   the hybrid model is about 1 order of magnitude higher than that of the BZ model, and the
   jet efficiency of the hybrid model is
   roughly consistent with that of the numerical MHD
   simulations for Kerr black holes\citep{hk06}.

   (2) The jet power dominates over the accretion power when the
   accretion rate is less than a critical value $\dot{m}_{\rm c}$,
   while the accretion power will be dominant when the accretion rate is larger than this critical
   value (Fig. 2), which is roughly consistent with that constrained from the
   observations of XRBs \citep*[e.g.,][]{fe03,mf06}.

   (3)  The dividing line of the Ledlow-Owen relation in $Q_{\rm jet}-M_{\rm BH}$ plane for
  the FR I/FR II dichotomy can be well reproduced by the hybrid jet model,
  provided that the ADAFs surrounding Kerr black holes are
  accreting at $\dot{m}\la 0.01$ in FR I sources(Fig. 3). The ionization luminosity
  corresponding to the dividing line plane also suggests the critical accretion rate to be $\sim 0.01$
  if the empirical relation between the photo-ionizing luminosity
  and the radio luminosity is adopted. These strongly imply that most FR
  I radio galaxies may have a different accretion mode from FR II sources.

\acknowledgements The authors are grateful to the referee for
his/her constructive suggestions on our paper. This work is partly
supported by the NSFC (grant 10773020, 10633010 and 10703009), and
the CAS (grant KJCX2-YW-T03). Q. W. W. thanks the postdoctoral
financial support from the Korean Astronomy and Space Science
Institute.

\clearpage
%\begin{figure}
%\epsscale{1.0} \plotone{f1.eps} \caption{The relation between the
%jet power $Q_{\rm jet}$ extracted from the underlying ADAF and BH
%spin $j$ for the BZ and hybrid models, respectively. The black solid
%lines and blue dashed lines denote $\alpha=0.3$ ($\beta=1$) and
%$\alpha=0.1$ ($\beta=5$ and $\delta=0.5$), respectively.  The red
%dotted lines denote $\alpha=0.3$ for the case of $\delta=0.1$. The
%BH mass $M_{\rm BH}=10^{8}\msun$ and accretion rate $\dot{m}=0.01$
%are adopted in the calculations. \label{fig1}}
%\end{figure}

%\begin{figure}
%\epsscale{1.0} \plotone{f2.eps} \caption{The relation between
%$Q_{\rm jet}/L_{\rm Edd}-\dot{m}$ (dashed line) and  $L_{\rm
%bol}/L_{\rm Edd}-\dot{m}$ (solid line) for BH spin parameters:
%$j=0.99,0.9,0.7,0$(from top to bottom), respectively. \label{fig2}}
%\end{figure}

%\begin{figure}
%\epsscale{1.0} \plotone{f3.eps} \caption{Jet power and BH mass
%dividing line between FR I and FR II radio galaxies for the
%\citet{lo96} sample (\emph{solid lines}; the jet power is calculated
%from the 151 MHz radio luminosity with two different correction
%factors $f=20$ (top) and $f=10$ (bottom), respectively). The dashed
%and dotted lines are the jet power of the hybrid model and BZ model
%extracted from the underlying ADAFs with $j=0.99,0.9,0.7$
%(\emph{from top to bottom}), respectively. The accretion rate
%$\dot{m}=0.01$ is adopted in all calculations, which is roughly
%consistent with that constrained from the ionization luminosity of
%the dividing line between FR I and FR II sources (see the text for
%more details).
% \label{fig3}}
%\end{figure}

\end{document}